\documentstyle[12pt]{article}
\newlength{\pubnumber} 

\catcode`\@=11
\@addtoreset{equation}{section}

\def\section{\@startsection{section}{1}{\z@}{3.5ex plus 1ex minus .2ex}
 {2.3ex plus .2ex}{\large\bf}}
\def\subsection{\@startsection{subsection}{2}{\z@}{2.3ex plus .2ex}
 {2.3ex plus .2ex}{\bf}}

\begin{document}

\begin{titlepage}
\samepage{
\rightline{CERN-TH/98-65}
\rightline{LPTHE-ORSAY 98/25}
\rightline{\tt hep-ph/9803466}
\rightline{March 1998}
\vfill
\begin{center}
   {\Large \bf Extra Spacetime Dimensions and Unification\\}
\vfill
   {\large
    Keith R. Dienes$^1$\footnote{
     E-mail address: keith.dienes@cern.ch},
    Emilian Dudas$^{1,2}$\footnote{
     E-mail address:  emilian.dudas@cern.ch},
     $\,$and$\,$
      Tony Gherghetta$^1$\footnote{
     E-mail address: tony.gherghetta@cern.ch}
    \\}
\vspace{.18in}
 {\it  $^1$ Theory Division, CERN, CH-1211 Geneva 23, Switzerland\\}
\vspace{.04in}
 {\it  $^2$ LPTHE, Univ.\ Paris-Sud, F-91405, Orsay Cedex, France\\}
\end{center}
\vfill
\begin{abstract}
  {\rm
     We study the effects of extra spacetime dimensions at
     intermediate mass scales, as expected in string theories with 
     large-radius compactifications, and focus 
     on the gauge and Yukawa couplings within the Minimal Supersymmetric
     Standard Model. We find that extra spacetime dimensions naturally lead
     to the appearance of grand unified theories at scales
     substantially below the usual GUT scale. 
     Furthermore, we show that extra spacetime dimensions provide 
     a natural mechanism for explaining the fermion mass hierarchy
     by permitting the Yukawa couplings 
     to receive power-law corrections. 
     We also discuss how proton-decay constraints may be
     addressed in this scenario, and suggest that proton-decay amplitudes
     may be exactly cancelled to all orders in perturbation theory 
     as a result of new Kaluza-Klein selection rules     
     corresponding to the extra spacetime dimensions.}
\end{abstract}
\vfill
\smallskip}
\end{titlepage}

\setcounter{footnote}{0}

\newcommand{\newc}{\newcommand}

\newc{\gsim}{\lower.7ex\hbox{$\;\stackrel{\textstyle>}{\sim}\;$}}
\newc{\lsim}{\lower.7ex\hbox{$\;\stackrel{\textstyle<}{\sim}\;$}}

\def\beq{\begin{equation}}
\def\eeq{\end{equation}}
\def\beqn{\begin{eqnarray}}
\def\eeqn{\end{eqnarray}}
\def\sosixteen{{$SO(16)\times SO(16)$}}
\def\e8{{$E_8\times E_8$}}
\def\V#1{{\bf V_{#1}}}
\def\half{{\textstyle{1\over 2}}}
\def\ttwo{{\vartheta_2}}
\def\tthree{{\vartheta_3}}
\def\tfour{{\vartheta_4}}
\def\ttwob{{\overline{\vartheta}_2}}
\def\tthreeb{{\overline{\vartheta}_3}}
\def\tfourb{{\overline{\vartheta}_4}}
\def\etainv{{\overline{\eta}}}
\def\Str{{{\rm Str}\,}}
\def\bone{{\bf 1}}
\def\chibar{{\overline{\chi}}}
\def\Jbar{{\overline{J}}}
\def\qbar{{\overline{q}}}
\def\calO{{\cal O}}
\def\calE{{\cal E}}
\def\calT{{\cal T}}
\def\calM{{\cal M}}
\def\calF{{\cal F}}
\def\calY{{\cal Y}}
\def\rep#1{{\bf {#1}}}
\def\ie{{\it i.e.}\/}
\def\eg{{\it e.g.}\/}
\def\eleven{{(11)}}
\def\ten{{(10)}}
\def\nine{{(9)}}
\def\Ip{{\rm I'}}
\def\oneprime{{I$'$}}
\hyphenation{su-per-sym-met-ric non-su-per-sym-met-ric}
\hyphenation{space-time-super-sym-met-ric}
\hyphenation{mod-u-lar mod-u-lar--in-var-i-ant}


\def\inbar{\,\vrule height1.5ex width.4pt depth0pt}

\def\IC{\relax\hbox{$\inbar\kern-.3em{\rm C}$}}
\def\IQ{\relax\hbox{$\inbar\kern-.3em{\rm Q}$}}
\def\IR{\relax{\rm I\kern-.18em R}}
 \font\cmss=cmss10 \font\cmsss=cmss10 at 7pt
\def\IZ{\relax\ifmmode\mathchoice
 {\hbox{\cmss Z\kern-.4em Z}}{\hbox{\cmss Z\kern-.4em Z}}
 {\lower.9pt\hbox{\cmsss Z\kern-.4em Z}}
 {\lower1.2pt\hbox{\cmsss Z\kern-.4em Z}}\else{\cmss Z\kern-.4em Z}\fi}

\def\NPB#1#2#3{{\it Nucl.\ Phys.}\/ {\bf B#1} (19#2) #3}
\def\PLB#1#2#3{{\it Phys.\ Lett.}\/ {\bf B#1} (19#2) #3}
\def\PRD#1#2#3{{\it Phys.\ Rev.}\/ {\bf D#1} (19#2) #3}
\def\PRL#1#2#3{{\it Phys.\ Rev.\ Lett.}\/ {\bf #1} (19#2) #3}
\def\PRT#1#2#3{{\it Phys.\ Rep.}\/ {\bf#1} (19#2) #3}
\def\CMP#1#2#3{{\it Commun.\ Math.\ Phys.}\/ {\bf#1} (19#2) #3}
\def\MODA#1#2#3{{\it Mod.\ Phys.\ Lett.}\/ {\bf A#1} (19#2) #3}
\def\IJMP#1#2#3{{\it Int.\ J.\ Mod.\ Phys.}\/ {\bf A#1} (19#2) #3}
\def\NUVC#1#2#3{{\it Nuovo Cimento}\/ {\bf #1A} (#2) #3}
\def\etal{{\it et al.\/}}

\long\def\@caption#1[#2]#3{\par\addcontentsline{\csname
  ext@#1\endcsname}{#1}{\protect\numberline{\csname
  the#1\endcsname}{\ignorespaces #2}}\begingroup
    \small
    \@parboxrestore
    \@makecaption{\csname fnum@#1\endcsname}{\ignorespaces #3}\par
  \endgroup}
\catcode`@=12

\input epsf


\section{Introduction}
\setcounter{footnote}{0}

The appearance of extra spacetime dimensions at high energy
scales is a generic feature of string theory. Typically these 
extra dimensions remain compactified at the Planck scale, but it is 
possible for new dimensions to have an effect below the Planck scale.
In particular, large-radius compactification schemes have recently 
been discussed in a number of theoretical and phenomenological 
contexts~\cite{largeradius}. Similarly, the effects of extra dimensions 
below the Planck scale have played a role in understanding the 
strong-coupling behaviour of string theory. In the Ho\v{r}ava-Witten 
scenario~\cite{HW}, the gravitational coupling feels the effect 
of an extra (bulk) dimension, while the gauge couplings, 
which are confined to the boundary, remain immune. 
A more phenomenological point of view, although still within the confines 
of string theory, has been considered in Ref.~\cite{antoniadis}, where 
the low-energy consequences of a single large (TeV-scale) dimension have 
been studied. Again the string models were constructed in such a way as 
to shield the gauge couplings from the extra dimensions. 
More recently, there has even been a field-theoretic
proposal for extra dimensions in the millimetre range~\cite{Dim}. 
Once again, the gauge couplings were shielded from the extra dimensions.

In this paper, by contrast, we will not shield the gauge and Yukawa 
couplings from the effects of the extra dimensions. Rather, we will focus
on the effects that these extra dimensions have on the gauge and Yukawa
couplings in the Minimal Supersymmetric Standard Model (MSSM), and discuss
how these effects may be exploited for phenomenological purposes.

The appearance of extra spacetime dimensions at an intermediate mass scale 
affects the gauge and Yukawa couplings in a nontrivial way. These effects
can be calculated in purely field-theoretic terms for large radii 
and there is no need 
to invoke string theory other than to provide the underlying motivation.
Formally, the appearance of an extra spacetime dimension is equivalent to the 
introduction of an infinite number of Kaluza-Klein states.
Of course, increasing the spacetime dimensionality in field theory makes the
divergences worse and leads to a loss of renormalisability. 
However, at any particular scale, we can assume that the contributions 
from the Kaluza-Klein states with masses larger than the scale of 
interest are decoupled from the theory, giving rise to an approximately 
renormalisable field theory. In this approximation we can calculate 
and trust the corrections to the gauge and Yukawa couplings. 
We will find that for a minimal set of Kaluza-Klein states, 
corresponding to a Kaluza-Klein tower for the gauge and Higgs bosons of 
the MSSM, the power-law corrections to the gauge couplings 
preserve gauge coupling unification, but lower the unification scale
considerably. This scale can naturally be interpreted as the fundamental 
mass scale of a further underlying theory (perhaps even a string theory).
Moreover, the finite power-law
corrections to the Yukawa couplings have the right sign and magnitude
to cancel the tree-level terms. This can help to explain the hierarchical 
structure of the fermion Yukawa couplings.

A detailed phenomenological analysis of the effects of extra spacetime 
dimensions will be presented in Ref.~\cite{ddg}. In this paper, we 
will concentrate on presenting the basic ideas and outline the calculation
of the effects of extra dimensions. We begin in Sect.~2 with a discussion 
of how the Kaluza-Klein states of the gauge and Higgs bosons affect the 
gauge couplings. Remarkably, the unification of the gauge couplings will 
remain intact, and the consequences for low-energy phenomenology will be
discussed. In Sect.~3 we consider the corresponding scenario for the 
Yukawa couplings and discuss the consequences for the fermion mass
hierarchy. Our conclusions and final comments will be presented in Sect.~4.

\section{Extra dimensions and gauge coupling unification}
\setcounter{footnote}{0}

We will assume that there exist $\delta\equiv D-4$ extra spacetime 
dimensions of 
fixed radius $R$, where $R^{-1}$ exceeds presently observable 
energy scales. Thus $\mu_0\equiv R^{-1}$ is the corresponding 
mass scale above which extra spacetime dimensions effectively 
appear. In principle, every particle state in the MSSM of mass $m_0$ 
can have an infinite tower of Kaluza-Klein states with masses
\beq
         m^2_n ~\equiv~ m^2_0~+~ \sum_{i=1}^\delta \,{n_i^2\over R^2}~,
\label{KKmasses}
\eeq
where each state exactly mirrors the zero-mode MSSM ground state 
and $n_i\in\IZ$ are the corresponding Kaluza-Klein  
excitation numbers. In the following we can safely neglect the zero 
mode mass $m_0$ in (\ref{KKmasses}) since $R^{-1}$ is presumed to exceed 
presently observable energy scales.  
However, in order to define consistent Kaluza-Klein masses in higher
dimensions we need to
introduce additional particles into the spectrum. A minimal extension
is to consider a Kaluza-Klein tower for the gauge and Higgs bosons and
assume that Kaluza-Klein excitations of the chiral fermions are absent. 
It is straightforward to identify all the states, since the massive 
Kaluza-Klein states fall into representations of $N=2$ supersymmetry.
Thus, at each Kaluza-Klein mass level $n$, the particle content of the 
MSSM is augmented 
by an $N=2$ vector supermultiplet for each gauge group, and an 
$N=2$ hypermultiplet for the two Higgs doublets. In component form we have 
\begin{equation}
\label{Tmdefn}
  V=\left(\begin{array}{cc} A_\mu^{(n)} & \phi^{(n)}\\
                            \lambda^{(n)} & \psi^{(n)}\end{array}\right)\,,
  \qquad
  H=\left(\begin{array}{cc} H_1^{(n)} & H_2^{(n)}\\
                            \psi_{H_1}^{(n)} & \psi_{H_2}^{(n)}
                            \end{array}\right)
\end{equation}
where all gauge indices have been suppressed.
In terms of four-dimensional $N=1$ multiplets, the $N=2$ vector 
supermultiplet corresponds to an $N=1$ vector multiplet and an $N=1$ chiral
multiplet in the adjoint representation of the gauge group. One of the
real scalar fields in the chiral multiplet becomes the longitudinal
component of the massive gauge boson, while the other real scalar field 
and the Weyl fermion remain in the spectrum at the massive level. 
The $N=2$ hypermultiplet represents the two massive Higgs doublets. 

The fact that the massive states fall into representations of 
$N=2$ supersymmetry is strictly true for $\delta=1,2$. For $\delta\geq 2$, 
the compact coordinates in string theory can be naturally
complexified; for example, the $\delta=6$ case implies the
existence of three complex ``planes''. In general orbifold models, gauge 
couplings can receive corrections~\cite{kaplunovsky} only if some 
orbifold element leaves a plane invariant;
they will then depend on the moduli (radii and angles) corresponding
to the invariant plane. There are no nontrivial orbifold elements which
simultaneously leave two planes invariant. Thus, without loss of
generality we can restrict to $\delta=1$ or 2 in the following.

Notice that in order to have consistent Kaluza-Klein gauge boson masses,
we had to introduce an $N=1$ chiral adjoint at each massive level. If there
were a Kaluza-Klein tower for the chiral fermions as well, then we
would have needed to introduce a set of mirror fermions in order to 
consistently define a Dirac mass in higher dimensions. While this is 
certainly possible,
we will only consider Kaluza-Klein excitations for the non-chiral
sector of the MSSM. (In string language, this is equivalent to assuming
that the non-chiral states arise in the twisted sectors of the theory.  
This ``minimal'' extension of the MSSM to higher dimensions is therefore 
similar to that considered in Ref.~\cite{antoniadis}.) Thus, below the scale 
$\mu_0$, our scenario consists 
of the regular MSSM states, with the gauge couplings running in the usual 
logarithmic fashion. At the scale $\mu_0$, the effects of extra 
spacetime dimensions compactified on a circle of radius $R$ become 
significant due to the appearance of 
massive gauge and Higgs bosons. For $\mu\gg\mu_0$, 
excitations of many Kaluza-Klein modes
become possible, and the contributions of these Kaluza-Klein states 
must be included in all physical calculations.

Let us now consider the effects of these Kaluza-Klein states on the 
gauge couplings. In ordinary four-dimensional field theory, 
the one-loop corrections to the gauge couplings $g_i$ are given by
\beq
         \alpha_i^{-1}(\mu) ~=~ 
         \alpha_i^{-1}(M_Z) ~-~ {b_i\over 2\pi}\,\ln {\mu\over M_Z}~,
\label{oneloopRGEfour}
\eeq
where $\alpha_i\equiv g_i^2/4\pi$, and the $b_i$ are
the MSSM one-loop $\beta$-function coefficients 
\beq
        (b_1,b_2,b_3)~=~ (33/5,1,-3)~.
\label{bdef}
\eeq
We will take the $Z$-mass $M_Z\equiv 91.17$ GeV as 
an arbitrary low-energy reference scale and at this scale 
(and within the $\overline{\rm MS}$ renormalisation group scheme),
the gauge couplings are given by
\beqn
        \alpha_Y^{-1}(M_Z)|_{\overline{\rm MS}} &\equiv&  98.29 \pm 0.13
                    \nonumber\\
        \alpha_2^{-1}(M_Z)|_{\overline{\rm MS}} &\equiv&  29.61 \pm 0.13
                    \nonumber\\
        \alpha_3^{-1}(M_Z)|_{\overline{\rm MS}} &\equiv&  8.5 \pm 0.5 ~.
\label{lowenergycouplingsa}
\eeqn
As is well-known, an extrapolation of these low-energy couplings 
according to (\ref{oneloopRGEfour}) leads to the celebrated unification 
relation
\beq
          \alpha_1(M_{\rm GUT}) ~=~ \alpha_2(M_{\rm GUT})
               ~=~ \alpha_3(M_{\rm GUT}) ~\approx ~ {1\over 24} 
\eeq
at the unification scale $M_{\rm GUT} ~\approx~ 2 \,\times\, 10^{16} 
~{\rm GeV}$.

Let us now consider the effects of the extra spacetime dimensions. The usual
one-loop corrections to the gauge couplings arise from the
vacuum polarisation diagram where zero-mode masses are included
in the loops. To compute the effects of extra dimensions we need 
to include the massive Kaluza-Klein states in the loops as well. 
While the full calculation for the MSSM 
can be done, for the sake of clarity we will first outline
an analogous but much simpler calculation involving the 
Kaluza-Klein states of a single Dirac fermion charged under 
a $U(1)$ gauge group.
We will then generalise the results to the full MSSM. 

For a single Dirac fermion with Kaluza-Klein excitations, the vacuum 
polarisation contribution is given by
\beq
   \Pi_{\mu\nu}(k^2)~=~ -\,
     \sum_{n_i=-\infty}^\infty\, g^2 \, 
    \int_0^\infty {d^4 q\over (2\pi)^4} \, {\rm Tr}\,\left(
    \gamma_\mu {1\over \not q - m_n } \gamma_\nu 
    {1\over \not k + \not q - m_n}  \right)
\label{firsteq}
\eeq
where we have used the notation
\beq
      \sum_{n_i= -\infty}^\infty ~\equiv ~ 
     \sum_{n_1= -\infty}^\infty 
     \sum_{n_2= -\infty}^\infty 
     \cdots
     \sum_{n_\delta= -\infty}^\infty 
\eeq
to represent a summation over all corresponding Kaluza-Klein excitations 
with masses $m_n^2$ given in (\ref{KKmasses}). Here
$m_0$ is the energy of the ground state, which we will henceforth
take to be zero for simplicity. The summation
over the Kaluza-Klein states can be performed with the aid of 
the Jacobi $\vartheta_3$ function,
\beq
  \vartheta_3(\tau) ~\equiv~ \sum_{n=-\infty}^\infty \,
  \exp(\pi i \tau n^2)~
\eeq
where $\tau$ is a complex parameter.
Using standard techniques to simplify the integral and introducing a 
Schwinger proper time parameter $t$ leads to the expression~\cite{ddg}
\beq
      \Pi(0) ~=~ { g^2\over 12\pi^2 } \,
             \int_0^\infty {dt\over t} \,
      \left\lbrace \vartheta_3\left( {it\over \pi R^2} \right)\right
      \rbrace^\delta~,
\label{intermed}
\eeq
where we have used the relation 
$\Pi_{\mu\nu}(k^2)=(k_\mu k_\nu -g_{\mu\nu}k^2)\,\Pi(k^2)$. The contribution
from the infinite tower of Kaluza-Klein states is contained in the Jacobi
$\vartheta_3$ function.

At this step, we must introduce our infrared and ultraviolet regulators,
along with their corresponding cutoffs. This can simply be done by 
introducing upper and lower cutoffs on the $t$-integration:
\beq
       \int_0^\infty \, dt ~\longrightarrow~
       \int_{r\Lambda^{-2}}^{r\mu_0^{-2}}  \, dt~.
\label{cutoff}
\eeq
In string theory, the issue of choosing suitable regulators 
has been considered in Ref.~\cite{Kiritsis}; however, this choice of 
regulator is sufficient for our purposes and will be discussed more fully 
in Ref.~\cite{ddg}.
Here $\Lambda$ is our ultraviolet cutoff, $\mu_0$ is our infrared cutoff,
and the numerical coefficient $r$ (which ultimately relates these cutoff 
parameters to the underlying physical {\it mass scales}) is defined as
\beq
           r~\equiv~ \pi \,(X_\delta)^{-2/\delta}
\eeq
where the numerical factor $X_\delta$ will be discussed below. The 
expression (\ref{intermed}) 
contains the complete contribution from the infinite tower of Kaluza-Klein 
states as a function of the 
radius scale $R$. In the $R\rightarrow 0$ limit, we find 
$\vartheta_3\rightarrow 1$ and thus 
we obtain the usual logarithmic corrections to the gauge coupling:
\beq
       \Pi(0)  ~=~ 
      -{g^2 \over 6\pi^2}\, \ln {\Lambda\over \mu_0}
      ~=~ {g^2 b\over 8\pi^2}\, \ln {\Lambda\over \mu_0}~.
\eeq
In the above expression we have identified $b=-4/3$ as the $\beta$-function 
coefficient of our single Dirac fermion.

Let us now generalise the above result to the MSSM where we have assumed that
only the non-chiral
states have Kaluza-Klein excitations. Since at each Kaluza-Klein massive 
level we have an $N=2$ vector multiplet and an $N=2$ hypermultiplet, the 
$\beta$-function coefficients multiplying the Jacobi $\vartheta_3$ 
function are
\beq
      (\tilde b_1,\tilde b_2,\tilde b_3) ~=~ (3/5, -3,-6)~.
\label{btilde}
\eeq
Thus the gauge couplings receive corrections of the form
\beq
       \alpha_i^{-1}(\Lambda) ~=~ \alpha_i^{-1}(\mu_0) ~-~
            {b_i-\tilde b_i\over 2\pi}\,\ln{\Lambda\over \mu_0} 
          ~-~ {\tilde b_i\over 4\pi}\,
             \int_{r\Lambda^{-2}}^{r\mu_0^{-2}} {dt\over t} \,
     \left\lbrace \vartheta_3\left( {it\over \pi R^2} \right) 
     \right\rbrace^\delta~.
\label{KKresult}
\eeq
The result (\ref{KKresult}) gives the exact corrections of the MSSM
gauge couplings in the presence of an infinite tower of Higgs and gauge-boson
Kaluza-Klein states associated with $\delta$ extra dimensions 
compactified on circles of radius $R$.  

In the limit $\Lambda R\gg 1$, the exact result (\ref{KKresult}) reduces 
to the form
\beq
\label{powerlaw}
       \alpha_i^{-1}(\Lambda) ~=~ \alpha_i^{-1}(\mu_0) ~-~
            {b_i-\tilde b_i\over 2\pi}\,\ln{\Lambda\over \mu_0} 
    ~-~ {\tilde b_i X_\delta \over 2\pi\delta}\,  \, 
      \left\lbrack \left({\Lambda\over \mu_0}\right)^\delta-1\right\rbrack~,
\eeq
where we have used the relation
\beq
      \vartheta_3 \left( {it \over \pi R^2} \right) 
            \approx~    R\sqrt{\pi\over t}~
\label{approxed}
\eeq
to perform the integral in (\ref{KKresult}). Note that in the 
$\Lambda R\gg1$ limit, (\ref{powerlaw}) agrees with the result 
of a full string calculation since the string winding states decouple and the 
effective field theory can be safely used.
The expression (\ref{powerlaw}) is exactly the result that one would have
obtained by performing all loop integrals in a $(\delta+4)$-dimensional 
spacetime. This can be seen by noting that 
in $D$ spacetime dimensions, the
gauge couplings $\tilde g_i$ accrue a classical mass dimension  
\beq
      \lbrack \tilde g_i \rbrack ~=~ 2-{D\over 2}~~~~\Longrightarrow~~~~
      \lbrack \tilde \alpha_i^{-1}\rbrack ~=~ D-4 ~=~\delta~.
\label{classdims}
\eeq
However, since our extra spacetime dimensions 
have a fixed radius $R$, the four- and $D$-dimensional gauge couplings 
are related to each other via
\beq
          \alpha_i~=~ R^{-\delta}\, \tilde\alpha_i~.
\label{Irelation}
\eeq
Thus, in the expression (\ref{powerlaw}), the power-law behaviour can be 
viewed as the ``classical scaling'' that we expect the gauge couplings 
to experience due to their enhanced classical mass dimensions when the
spacetime dimension is bigger than four.

The fact that the infinite Kaluza-Klein summation is equivalent
to performing loop integrals in flat $D$-dimensional spacetime highlights 
the nonrenormalisable nature of the theory. Physical parameters now
depend on the cutoff scale $\Lambda$ and the normalisation of this scale
is parametrised by the coefficient $X_\delta$ that appears in (\ref{powerlaw}).
Na\"\i vely, we would expect $X_\delta$ to essentially be a 
$D$-dimensional phase space factor resulting from a $D$-dimensional 
loop integral.
However, this would be an incorrect assessment because one could alternatively
interpret this same factor as a normalisation for the cutoffs
that appear in the $D$-dimensional integrations. This makes it obvious 
that within the context of our nonrenormalisable field theory,
the coefficient $X_\delta$ is essentially cutoff- and regulator-dependent.
It turns out~\cite{ddg} that given the regulator chosen in (\ref{cutoff}), 
the appropriate value of $X_\delta$ is:
\beq
     X_\delta~=~ {\pi^{\delta/2}\over \Gamma(1+\delta/2)} ~=~
       {2 \pi^{\delta/2} \over \delta \Gamma(\delta/2)}~
\label{Xdef}
\eeq
where $\Gamma$ is the Euler gamma function.
Thus, $X_0=1$ (as expected), while $X_1=2$, $X_2= \pi$, $X_3=4\pi/3$, 
and so forth. Given this choice for $X_\delta$, it is then 
legitimate~\cite{ddg} to interpret $\Lambda$ as the 
mass scale for new physics beyond our effective nonrenormalisable theory.

We have seen that the contribution from the infinite tower of Kaluza-Klein
states is given by the exact expression (\ref{KKresult}), and 
reduces to the expression (\ref{powerlaw}) in the limit $\Lambda R\gg 1$. 
In practice, the exact expression (\ref{KKresult}) can be approximated
as follows: at any scale below $\mu_0$, we can replace 
(\ref{KKresult}) with the usual 
logarithmic running (\ref{oneloopRGEfour}), while for any scale above 
$\mu_0$ we can use the expression (\ref{powerlaw}). If we now match
these two solutions at the scale $\mu_0$, this yields our final result
valid for all $\Lambda\geq \mu_0$:
\beq
     \alpha_i^{-1}(\Lambda) ~=~ 
      \alpha_i^{-1}(M_Z)   
      ~-~ {b_i \over 2\pi} \,\ln{\Lambda\over M_Z} 
      ~+~ {\tilde b_i \over 2\pi} \,\ln{\Lambda\over \mu_0} 
           ~-~ {\tilde b_i X_\delta\over 2\pi \delta} \,\left\lbrack
                 \left({\Lambda\over \mu_0}\right)^\delta -1\right\rbrack~.
\label{newsoln}
\eeq
It is shown in Ref.~\cite{ddg} that this is an excellent approximation
to the full result given in (\ref{KKresult}).

We can see from (\ref{newsoln}) that the dependence of the gauge couplings 
on the scale $\Lambda$ drastically changes the normal one-loop
running of the gauge couplings. Remarkably, however, there always exists
a value of $\Lambda$ such that the gauge couplings continue to unify! Moreover,
this property is robust and does not depend on the number $\delta$ 
of extra spacetime dimensions or where the scale $\mu_0$ of new dimensions 
appears.

\begin{figure}
\centerline{ \epsfxsize 3.25 truein \epsfbox {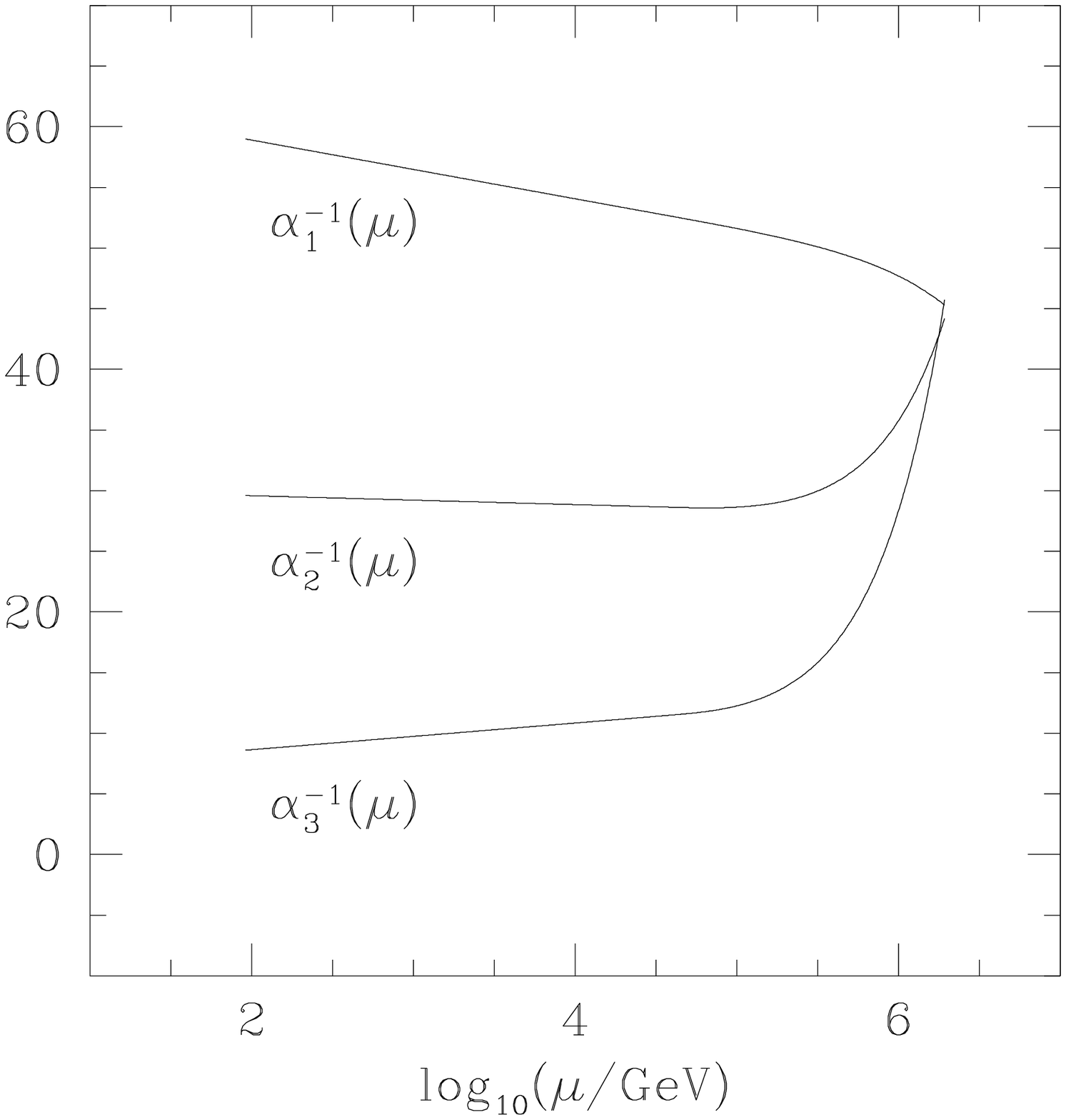}
             \epsfxsize 3.25 truein \epsfbox {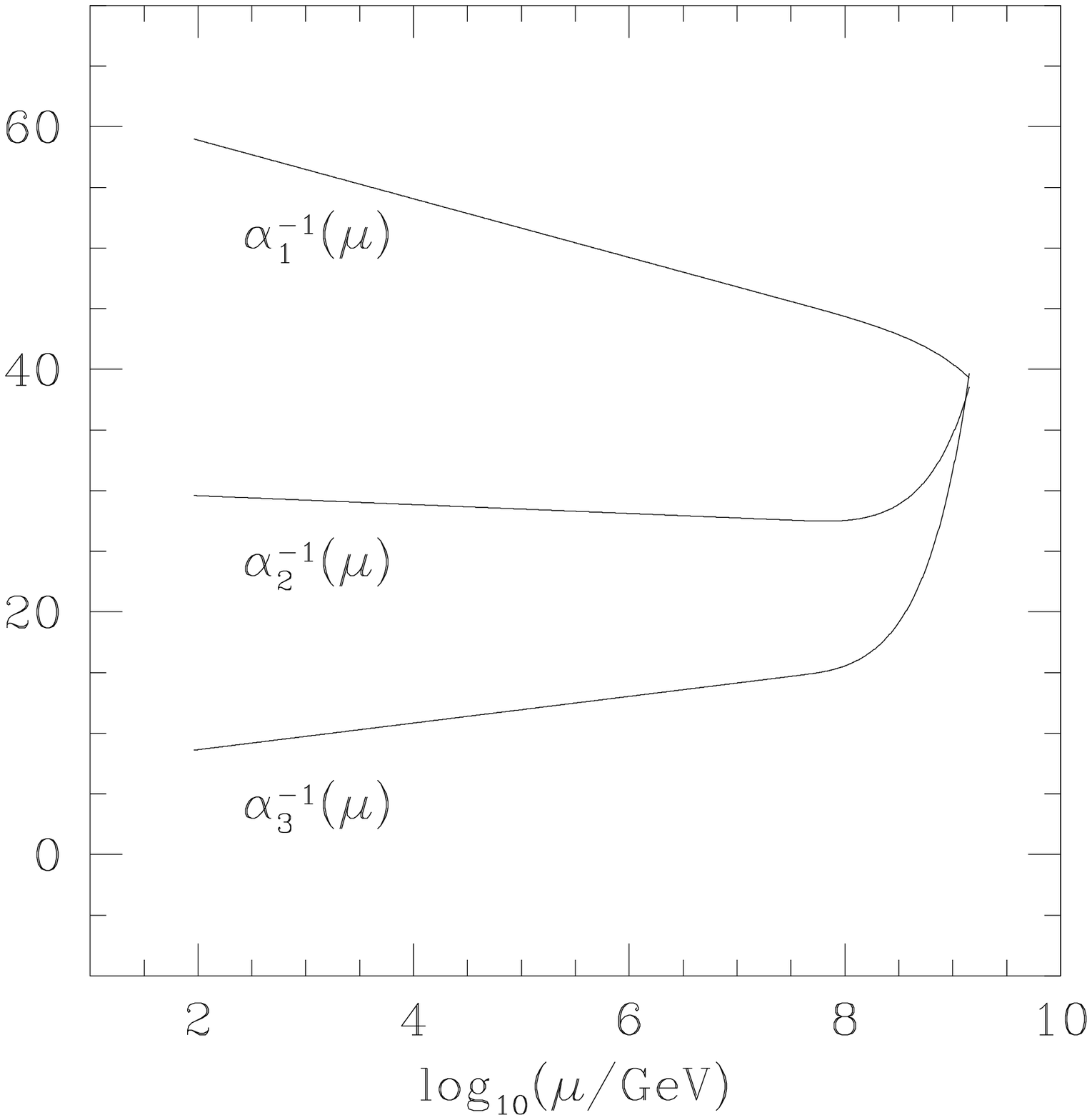}}
\centerline{ \epsfxsize 3.25 truein \epsfbox {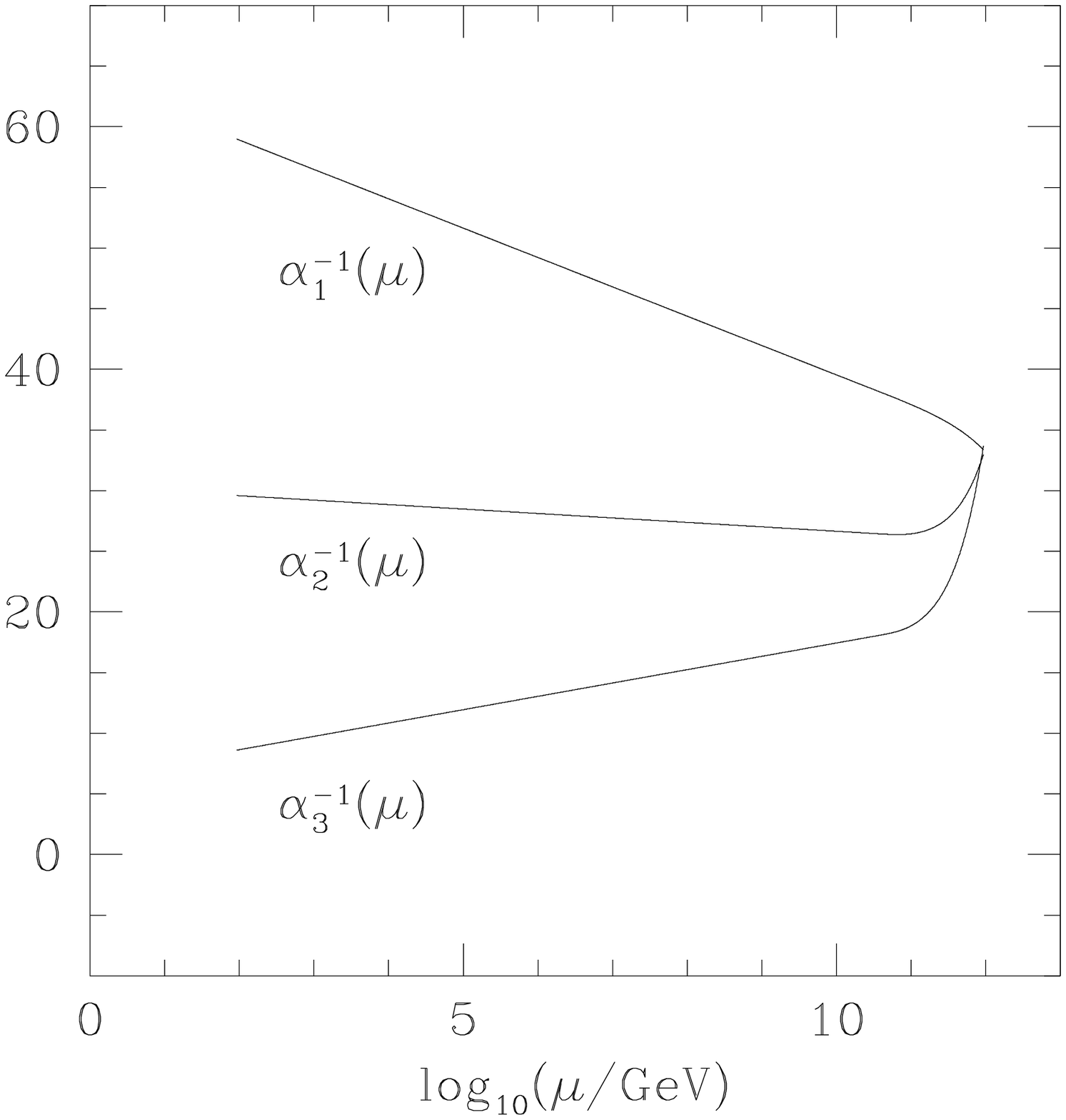}
             \epsfxsize 3.25 truein \epsfbox {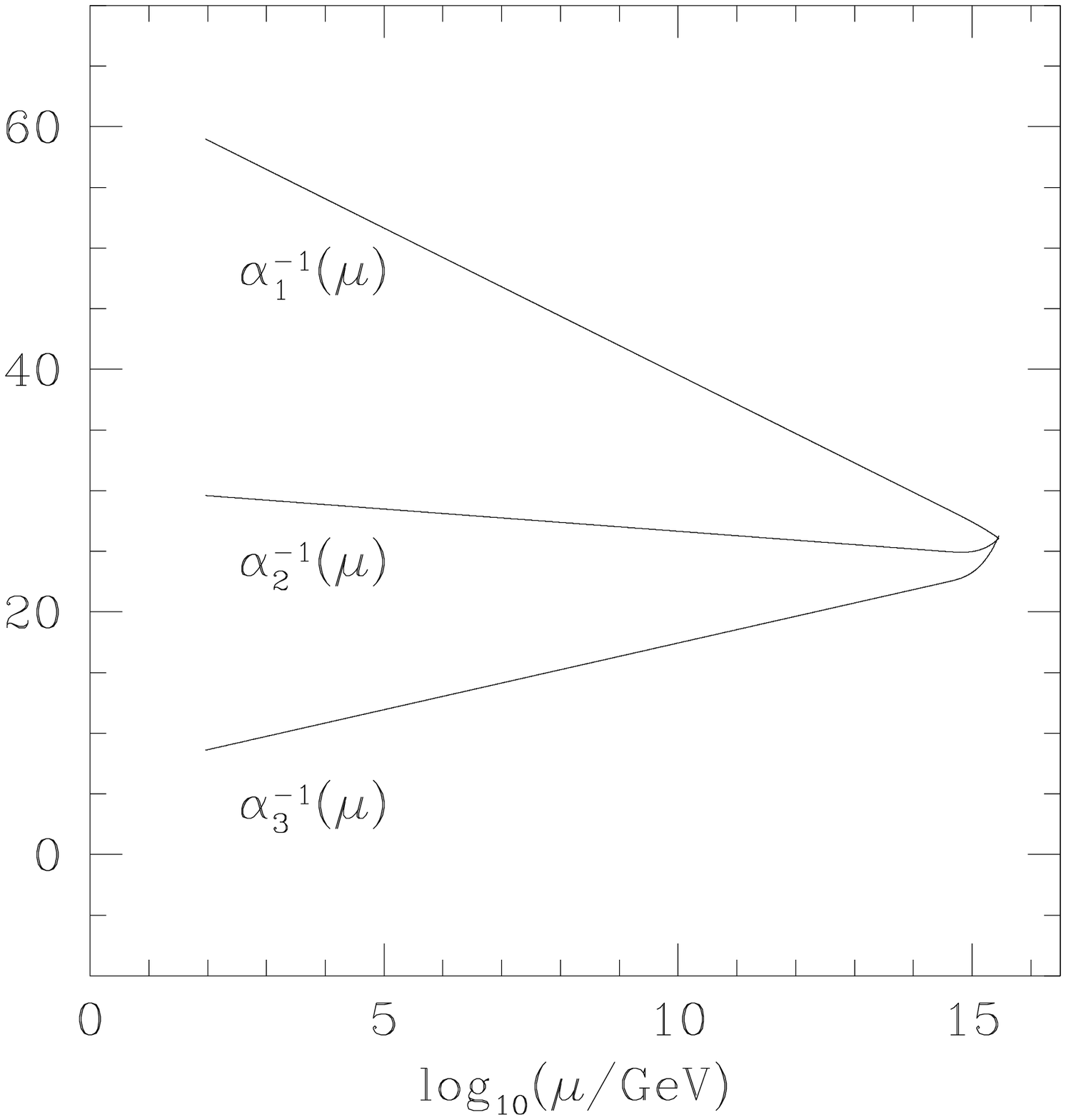}}
\caption{Unification of gauge couplings in the presence of
     extra spacetime dimensions.
     We consider four representative cases:  
          $\mu_0 =  10^{5}$ GeV (top left),
          $\mu_0 =  10^{8}$ GeV (top right),
          $\mu_0 =  10^{11}$ GeV (bottom left),  and
          $\mu_0 =  10^{15}$ GeV (bottom right).
      In each case we have taken $\delta=1$.  }
\label{unifII}
\end{figure}

This unification is illustrated in Fig.~\ref{unifII}, where we have
shown four representative cases. For $\mu\leq \mu_0$, we are plotting 
the usual running of the four-dimensional gauge couplings. For $\mu\geq 
\mu_0$, however, we are treating $\mu$ as the cutoff $\Lambda$ and 
plotting the values of the corrected gauge couplings as functions of 
this cutoff. We see that above the scale $\mu_0$, the appearance of 
extra spacetime dimensions
accelerates the ``running'' of the gauge couplings, due to the power-law 
corrections, and remarkably they continue to unify. This unification property
is directly related to the $\beta$-function coefficients (\ref{btilde}) 
of the $N=2$ matter content of the theory at each Kaluza-Klein mass level. It 
is noteworthy to point out that had we instead included mirror fermions 
and constructed Kaluza-Klein states for the chiral fermions, 
the $\beta$-function coefficients $\tilde b_2$ and $\tilde b_3$ would have 
changed signs and magnitudes, causing these couplings to become strong 
very quickly.

\begin{figure}[ht]
\centerline{ \epsfxsize 4.0 truein \epsfbox {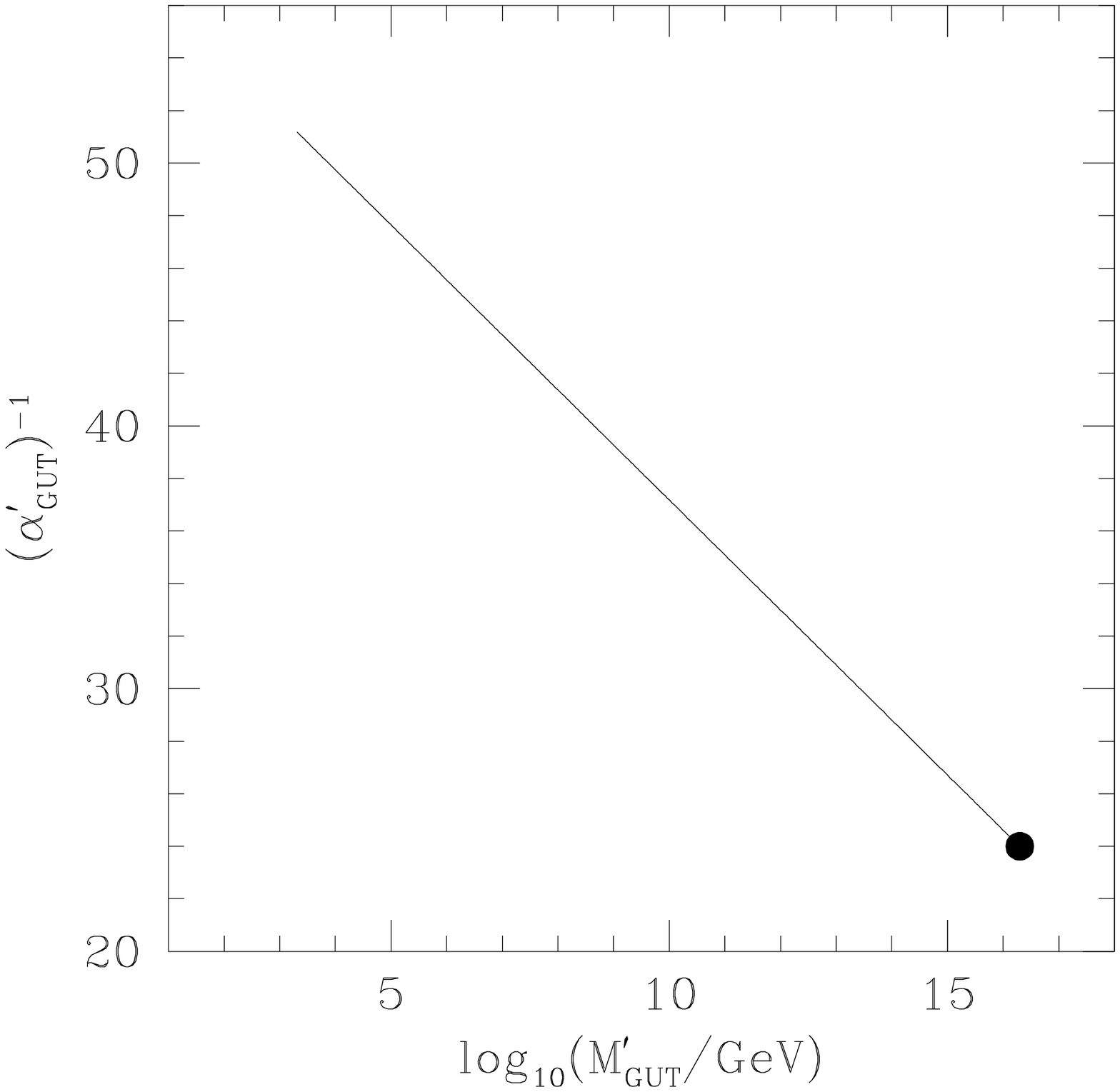}}
\caption{The unified coupling $(\alpha'_{\rm GUT})^{-1}$ as a function
       of the unification scale $M'_{\rm GUT}$.
       This curve is independent of the number of extra spacetime dimensions.
       The limit of the usual four-dimensional MSSM is indicated with a dot.}
\label{gaugethree}
\end{figure}

In our case, however, the $\beta$-functions have the correct sign and 
magnitude such that the gauge couplings unify at a value 
$\alpha_{\rm GUT}^\prime$
that remains weak. In fact, $\alpha_{\rm GUT}^\prime$ is even weaker
than the value $\alpha_{\rm GUT}$ obtained in the usual scenario. This 
behaviour is plotted in Fig.~\ref{gaugethree}. 
Thus, we see that our scenario\footnote{
     We are tempted to abbreviate
     this {\bf D}-{\bf D}\/imensional {\bf G}\/UT scenario as
     the DDG scenario, but modesty prevents us from doing so!  }
naturally predicts the emergence of
a $D$-dimensional GUT at the new lower scale $M'_{\rm GUT}$ for which 
the unified gauge
coupling is always more perturbative than in the MSSM!

The extent to which one can realistically lower the unification scale 
$M_{\rm GUT}^\prime$ below the usual value 
$M_{\rm GUT}\approx 2\times 10^{16}$ GeV
depends upon the crucial question of proton decay.
In the usual grand unification scenario, proton decay is effectively mediated 
at low energies by dimension-six terms in the Lagrangian which are
suppressed by inverse powers of the unification scale. In supersymmetric
models, dimension-five operators may also be relevant but can be suppressed. 
Even though
lowering the unification scale makes the proton-decay problem worse, our 
scenario does have a compensating factor since the unified gauge coupling 
$\alpha_{\rm GUT}^\prime$ becomes weaker. 
Thus, assuming that dimension-six terms are the dominant effective 
proton-decay operators, we find that the amplitude for proton decay 
in our scenario will only be enhanced by a factor
\beq
     {\alpha_{\rm GUT}^\prime \over \alpha_{\rm GUT}}
     \left( M_{\rm GUT} \over M'_{\rm GUT}\right)^2 ~.
\label{badfactor}
\eeq
Furthermore, we have slightly more freedom in lowering the unification scale
because the usual value $M_{\rm GUT}\approx 2\times10^{16}$ GeV oversatisfies
the experimental bounds. Since $M_{\rm GUT}^\prime$ depends on $\mu_0$ and
$\delta$, we can use the experimental bounds on the 
proton lifetime to derive a lower bound on $\mu_0$ for any $\delta$.
Specifically, we obtain $\mu_0\gsim 1\times 10^{14}$ GeV for $\delta=1$,
and $\mu_0\gsim 3\times 10^{14}$ GeV for $\delta=2$.
Thus, as long as the scale $\mu_0$ of the extra dimensions  
is sufficiently large, the usual proton-decay 
bounds can be satisfied in each case.

While our scenario for extra dimensions can be 
imposed at energy scales that are close to the usual unification scale
near $10^{16}$ GeV, we stress that it might nevertheless be possible 
to tolerate lower energy scales by finding an intrinsically 
{\it higher-dimensional} solution to the usual proton-decay problem. 
In particular, if we impose Kaluza-Klein selection
rules on the baryon number-violating couplings, then we could suppress
proton decay. For example, if the $X$ boson and coloured Higgs triplets
in an $SU(5)$ GUT had only odd
Kaluza-Klein excitations, then they could not couple at tree-level to the 
usual MSSM states (which have $n=0$). Thus, all perturbative proton-decay 
diagrams would generally vanish, and general nonrenormalisable operators
would have additional suppression due to the presence of the large radius.
Alternatively, we can avoid the proton-decay problem altogether by 
shifting the string scale to $M'_{\rm GUT}$ (e.g., via the 
Witten scenario~\cite{HW2}), and by recalling the fact (see 
Ref.~\cite{review} for a review) that string theory is generally
consistent with gauge coupling unification regardless of whether any GUT
gauge group appears. Indeed, one could even imagine shifting the string
scale all the way down to the electroweak scale~\cite{lykken}.
A detailed discussion of these mechanisms will be presented in 
Ref.~\cite{ddg}.

\section{Extra dimensions and the fermion mass hierarchy}
\setcounter{footnote}{0}

Given that extra spacetime dimensions induce power-law corrections 
for the gauge 
couplings, it is natural to ask whether the fermion mass
hierarchy might also be explained in our scenario. Unlike the 
usual logarithmic corrections in four-dimensional field theory,
the power-law corrections that arise from extra dimensions can dramatically 
affect the fermion Yukawa couplings. 

Let us first recall
how the Yukawa couplings $y_F$ (with $F=e,\mu,\tau,u,d,s,c,b,t$)
run within the usual four-dimensional MSSM. If we define 
$\alpha_F\equiv y_F^2/4\pi$ in analogy with the gauge couplings
$\alpha_i$, then the Yukawa coupling one-loop RGE's in the MSSM have the form
\beq
   {d\over d \ln \mu} \, \alpha^{-1}_F(\mu) ~=~ -{ b_F(\mu) \over 2\pi}~.
\label{yukrunning}
\eeq
Indeed, the only difference relative to the gauge couplings is that
the one-loop $\beta$-function ``coefficients'' $b_F(\mu)$ are not constants,
but instead depend on the scale $\mu$, since there are now
many different couplings which can contribute to the one-loop $\beta$-function.
For example, within the usual MSSM, $b_t(\mu)$ is given by:
\beq
   b_t ~\equiv~  6 ~+~ {1\over \alpha_t}\, \left( \alpha_b + 3\alpha_u 
   + 3\alpha_c -{ 16\over 3}\, \alpha_3 - 3 \alpha_2 - { 13 \over 15}\, 
   \alpha_1\right)~,
\eeq
and each of the other $b_F(\mu)$ has a similar form.

Let us now consider how the evolution of Yukawa couplings is modified in the
presence of extra spacetime dimensions. Just as for the gauge couplings,
we shall assume that a certain number $\delta$ of
extra spacetime dimensions appear at an energy scale $\mu_0\equiv R^{-1}$.
Below the scale $\mu_0$, the Yukawa couplings run according
to (\ref{yukrunning}). Note that the running implied by (\ref{yukrunning})
is solely a property of the four-dimensional renormalisable field theory.
Above the scale $\mu_0$, where we have a nonrenormalisable theory, 
the Yukawa couplings instead receive finite one-loop corrections whose 
magnitudes depend upon the cutoff scale $\Lambda$. 
By comparison with
our prior results for the gauge couplings, the extra dimensions will induce
corrections for the Yukawa couplings of the form
\beq
  \alpha_F^{-1}(\Lambda) ~=~ 
  \alpha_F^{-1}(\mu_0) - {b_F(\mu_0)-\tilde b_F(\mu_0)\over 2\pi}
  \ln {\Lambda\over \mu_0} ~-~  {\tilde b_F(\mu_0)\over 2\pi} 
  {X_\delta\over \delta}\, 
  \left\lbrack \left( {\Lambda\over\mu_0}\right)^\delta-1\right\rbrack~
\label{yuksoln}
\eeq
where we have matched onto the usual logarithmic running at $\Lambda=\mu_0$
and $\tilde b_F$ represent the contributions from the states that have
Kaluza-Klein excitations. The form of the above solution is 
qualitatively very similar to that for the gauge couplings. The only 
difference is that since the $\beta$-function 
coefficients for the Yukawa couplings are not pure numbers, the 
$\beta$-functions must be evaluated at the fixed scale $\mu_0$
at which we are evaluating our fixed nonrenormalisable one-loop corrections. 
Moreover, $X_\delta$ takes the same values as for the gauge couplings. 
These issues will be discussed more fully in Ref.~\cite{ddg}.

The computation of the coefficients $\tilde b_F$ proceeds in the usual way
by considering the one-loop anomalous dimensions of the fermions and
Higgs fields, with massive Kaluza-Klein states present in the loops. 
In our scenario, there are no contributions from the anomalous dimensions
of the Higgs fields because the non-chiral Kaluza-Klein excitations 
fall into $N=2$ representations. This is consistent with the fact that 
for $N=2$ hypermultiplets there is no wavefunction renormalisation. 
Thus the only contribution can come from the anomalous dimensions of the 
fermions. 

Na\"\i vely one would expect that in the computation of the anomalous 
dimensions of the fermions, massive Kaluza-Klein states would not 
be present in the loops due to Kaluza-Klein momentum conservation.
However, since the chiral fermions do not have Kaluza-Klein excitations 
in our scenario, the interactions between the fermions and the
gauge-boson or Higgs Kaluza-Klein states occur only at orbifold
fixed points where
there is no Kaluza-Klein momentum conservation. Thus, in the minimal
scenario, power-law corrections occur only for the fermion wavefunction
renormalisation factors $Z_F$ and $Z_{\overline F}$ where
\begin{equation}
  \alpha_F(\Lambda)~=~Z_H^{-1} Z_F^{-1} Z_{\overline F}^{-1}\,\alpha_F(\mu_0) 
\label{Texactalp}
\end{equation}
and $Z_H$ is the Higgs field wavefunction renormalisation factor.
Specifically we have
\begin{equation}
  Z_F~=~1-2~\tilde\gamma_F(\mu_0)\,{X_\delta\over \delta}\, 
  \left\lbrack \left( {\Lambda\over\mu_0}\right)^\delta-1\right\rbrack~
\label{TZpl}
\end{equation}
and similarly for $Z_{\overline F}$, where $\tilde\gamma_F(\mu_0)$ is the 
anomalous dimension of the fermion $F$ evaluated at the scale $\mu_0$. 
The wavefunction renormalisation 
factor $Z_H$ receives only logarithmic corrections. If we linearise
(\ref{Texactalp}) then we obtain (\ref{yuksoln}) with $\tilde b_F(\mu_0)=
4\pi(\tilde\gamma_F(\mu_0)+\tilde\gamma_{\overline F}(\mu_0))$. 
In the MSSM, the 
anomalous dimensions of the fermions are almost always dominated by the gauge
couplings and we generically find that $\tilde\gamma_F(\mu_0) <0$. Thus, 
for large $\Lambda$, we obtain $\alpha_F(\Lambda)\rightarrow 0$, and 
the hierarchy between Yukawa couplings is affected only by factors 
of order one.

In order to obtain $\tilde\gamma(\mu_0) >0$ we need to increase the strength
of the non-gauge couplings. This can be done by introducing a 
four-dimensional chiral singlet field $S$ with a superpotential 
term $W_S=\lambda_S H_1 H_2 S$. Since
at the massive level there is no Kaluza-Klein momentum conservation
for a coupling of this form, $Z_{H_i}$ will receive power-law corrections 
of the form (\ref{TZpl}) where $\tilde\gamma_{H_i}(\mu_0)=\alpha_S/4\pi
\equiv \lambda_S^2/16\pi^2$.
Thus, we see that the $\tilde\gamma_{H_i}$ are always positive and 
universal.

\begin{figure}[ht]
\centerline{ \epsfxsize 4.0 truein \epsfbox {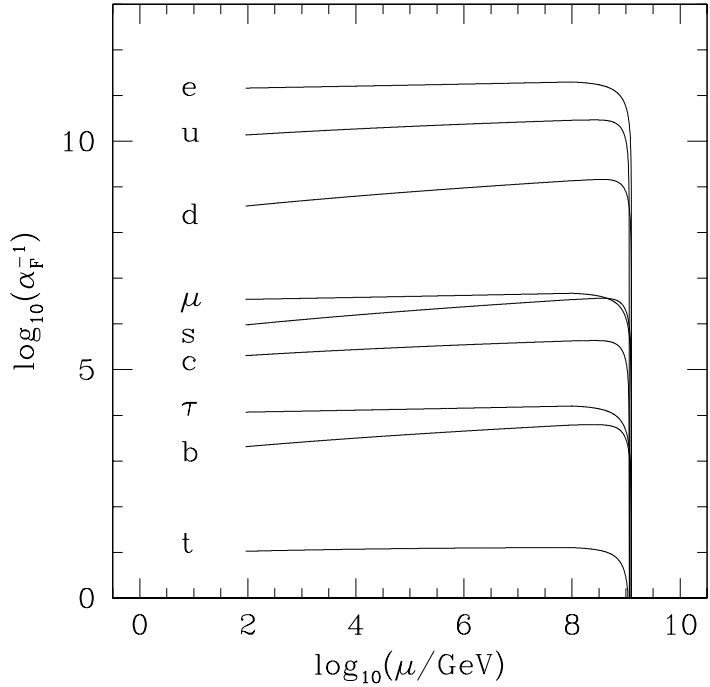}}
\caption{The evolution of the Yukawa couplings 
     $\alpha_F^{-1}\equiv 4\pi / y_F^2$
     within the MSSM, assuming the presence of a single extra 
     dimension at $\mu_0=10^{8}$ GeV.  
     We have taken $m_t=180$ GeV and $\tan\beta=3$ as a 
     representative case. Note that we are plotting the Yukawa 
     couplings on a {\it logarithmic}\/ scale
     in order to display them all simultaneously. It is evident that all of
     the Yukawa couplings approach a common Landau pole which precisely 
     agrees with the scale at which the gauge couplings unify.}
\label{yukplot1}
\end{figure}

An immediate consequence of this fact is that the one-loop
power-law corrections to $Z_{H_i}$ have the proper
magnitudes and signs to bring all of the Yukawa couplings simultaneously
to a common Landau pole scale. This can be explicitly seen in 
Fig.~\ref{yukplot1} where we have plotted the solution (\ref{Texactalp}) 
using the universal anomalous dimension coefficients $\tilde\gamma_{H_i}
(\mu_0)$ for $\alpha_S\simeq 1/4$. It is clear 
that above the scale $\mu_0$, the power-law term coming from the 
Kaluza-Klein states dominates the evolution, and the Yukawa couplings
tend towards a common large Yukawa coupling (\eg, towards a common 
Landau pole defined by the equation $Z_{H_i}=0$).
Note that because the power-law corrections for each fermion 
are not coupled to those of the other fermions, as would have 
been the case for the usual renormalisation group equations,
each fermion {\it independently} tends towards a Landau pole. Moreover,
for appropriate values of the coupling $\lambda_S$, this Yukawa 
``unification'' scale can naturally be associated
with the scale $M'_{\rm GUT}$ at which the gauge couplings unify.

Even though the Yukawa couplings tend towards a common Landau pole,
there still exists a hierarchy between any finite value of the Yukawa 
couplings if we only consider the power-law corrections. 
This stems from the universal coefficients $\tilde \gamma_H$.
However, the effect of the logarithmic terms in 
$Z_H$ is to slightly shift the position of the Landau pole
between the up- and down-type fermions. This is because $Z_{H_1}$ 
and $Z_{H_2}$ 
receive different logarithmic corrections. Thus, it is possible for 
up-type and down-type Yukawa couplings to pairwise cross and thereby 
completely eliminate the hierarchy between pairs of Yukawa couplings. 
It is also clear that the flavour-dependent 
corrections in $Z_F$ and $Z_{\overline F}$ do not affect the position of 
the Landau pole since this is controlled by $Z_H$. Therefore,
the superpotential $W_S$ cannot affect the hierarchy of the 
fermions within the up- or down-type sector.

In order to simultaneously reduce the
hierarchy for {\it all} fermions we need to introduce a 
flavour-dependent coupling. This can be simply achieved 
by introducing an MSSM singlet field $\Phi$ which couples to the
Yukawa-coupling term with a generic superpotential of the form
\begin{equation}
\label{TFNspot}
      W~=~\hat y_F \Phi^{n_F} \overline{F} F H
\end{equation}
where $\hat y_F$ is a dimensionful coupling and 
$\Phi$ has an associated Kaluza-Klein tower of massive states. 
Although this is reminiscent of the Froggatt-Nielsen scenario~\cite{FN},
we shall see that its implementation is different.
Again, since only a Dirac 
fermion mass can be consistently defined in higher dimensions, one
needs to introduce a conjugate superfield $\overline{\Phi}$ (with no couplings
to the MSSM fermions) and arrange both superfields into an $N=2$ 
hypermultiplet. The effect of $\Phi$ is to change the 
exponent of the power-law dependence for the $\beta$-function 
coefficients of the Yukawa couplings in a flavour-dependent way.

To see how this works, let us consider a single Yukawa coupling.
Using dimensional analysis we can form an effective dimensionless
Yukawa coupling $Y_F(\mu)\equiv\Lambda^{n_F} \hat y_F(\mu)$.
At the scale $\mu_0$, the physical Yukawa coupling is $y_F(\mu_0)\equiv
\hat y_F \langle\Phi\rangle^{n_F}=Y_F(\mu_0)(\mu_0/\Lambda)^{n_F}$, 
where we have assumed that $\Phi$ decouples below the scale $\mu_0$. 
In order to see the
effects of the higher dimensions we can compute the power-law corrections
arising from the superpotential (\ref{TFNspot}). The dominant power-law 
term comes from either $Z_F$ or $Z_{\overline F}$, and for 
a single effective dimensionless Yukawa coupling we obtain
\begin{equation}
  \alpha_F^{-1}(\Lambda) ~\simeq~ \alpha_F^{-1}(\mu_0) ~-~ 
  c_F \left\lbrack
  \left({\Lambda\over \mu_0}\right)^{(n_F+1)\delta} -1 \right\rbrack
\label{FNyukcorr}
\end{equation}
where $\alpha_F^{-1}\equiv 4\pi/Y_F^2$ and $c_F>0$ is a flavour-dependent 
constant. 
The power-law term comes from the $n_F$ copies of the $\Phi$ Kaluza-Klein 
states in $\delta$ extra dimensions and quickly drives the Yukawa coupling to
a Landau pole. If the Yukawa coupling is close to its Landau
pole scale, then $1/Y_F(\Lambda)\rightarrow 0$ and we obtain $Y_F^2
(\mu_0)\sim (\mu_0/\Lambda)^{(n_F+1)\delta}$. Thus, $Y_F(\mu_0)$ receives
an additional suppression from the extra $\delta$ dimensions,
which arises from the Yukawa coupling being near a Landau pole. 
Including this extra suppression yields $y_F^2(\mu_0)\sim 
(\mu_0/\Lambda)^{\Delta_F}$, 
where $\Delta_F\equiv \delta+n_F(2+\delta)$.
Thus, the effect of the extra spacetime dimensions is to increase
the exponent of the power-law corrections by an amount 
$(n_F+1)\delta$. In this way
the hierarchy of the Yukawa couplings can be explained without the
large values of $n_F$ that are needed in the usual Froggatt-Nielsen scenario. 
A complete analysis including all Yukawa couplings 
will be presented in Ref.~\cite{ddg}. 

\section{Conclusions and future prospects}
\setcounter{footnote}{0}

The appearance of extra spacetime dimensions at high energy scales
can have dramatic consequences on low-energy parameters. In particular,
we have seen that the gauge and Yukawa couplings receive power-law 
corrections which arise from an infinite tower of Kaluza-Klein states.
Alternatively, these power-law corrections may be interpreted 
as the classical scaling of dimensionful
gauge and Yukawa couplings. Remarkably, in our scenario, we find that
the gauge couplings continue to unify independently of the number of 
extra spacetime dimensions or the scale at which they are introduced. 
This leads to a $D$-dimensional GUT scenario in which the
value of the unified gauge coupling is even more perturbative than in the MSSM.
Our scenario may be safely implemented near scales of $10^{14}$ GeV, where
proton-decay constraints are satisfied. However, there exists the
possibility of invoking Kaluza-Klein selection rules to forbid proton-decay 
processes altogether. This is an inherently
higher-dimensional solution to the proton-decay problem and may even allow
the appearance of new dimensions near the TeV scale.
Furthermore, the Yukawa couplings receive power-law corrections of the right 
sign and magnitude to substantially ameliorate the 
Yukawa coupling hierarchy. Indeed, in our scenario, the Yukawa couplings 
all tend to unify at a scale which can be made to coincide with the scale of 
gauge coupling unification.

Our scenario clearly raises a number of intriguing questions.
First, in this paper we have merely presented a general 
scenario by which the appearance of extra spacetime dimensions can preserve
gauge coupling unification and also simultaneously ameliorate 
the fermion mass hierarchy.  However, it would be interesting to 
construct an explicit model in which our mechanism is realised, and 
in which proton decay is suppressed as a result of Kaluza-Klein selection 
rules. This could be done either through field theory, or through an
explicit string construction making use of large-radius compactifications
and associating the string scale with our unification scale.
It may also be possible to realise such constructions via suitable
brane configurations.
 
Another closely related issue concerns the unification of the
gauge couplings with the gravitational coupling.  It is clear that
the extra spacetime dimensions will also accelerate the 
``running'' of the gravitational coupling, in much the same way as
occurs in the Ho\v{r}ava-Witten scenario~\cite{HW}.
It is therefore possible that extra large spacetime dimensions
can induce a complete unification of gauge and
gravitational couplings --- all without making recourse to 
the strong-coupling dynamics of string theory.
 
Finally, we remark that it may not even be necessary to assume
the existence of supersymmetry in order for our scenario to achieve
gauge coupling unification.  This issue will be discussed further
in Ref.~\cite{ddg}. Moreover, supersymmetry is not as essential   
for solving the gauge hierarchy problem as it is within the MSSM 
if new spacetime dimensions populate  
the desert between the electroweak scale and the usual GUT scale.
Thus, our scenario may make it possible to 
achieve gauge coupling unification, Yukawa unification,
and also stabilise the Higgs mass, all without supersymmetry.
 
Thus, we see that that extra large spacetime dimensions are
naturally consistent with an intermediate-scale grand unified theory,
and may also ultimately help to explain the fermion mass hierarchy.
Moreover, extra large spacetime dimensions are a natural
mechanism whereby the phenomenological predictions of
string theory might be shifted downwards in energy scale so that
they might be more directly observed.
These issues, as well as other phenomenological consequences
of our scenario, will be discussed further in Ref.~\cite{ddg}.
 
\bigskip
\medskip
\leftline{\large\bf Acknowledgments}
\medskip

We wish to thank M.~Einhorn, G.~Kane, C.~Kounnas, J.~Pati and S.-H.H.~Tye 
for useful discussions. We are especially grateful to R.~Rattazzi 
for important comments on an earlier version of this paper.
KRD also wishes to thank the Department of Physics at the University 
of Michigan for hospitality where part of this work was done. 

\vfill\eject
\bigskip
\medskip

\bibliographystyle{unsrt}

\end{document}